\documentclass[%
 reprint, nofootinbib,
 amsmath,amssymb,
 aps,
    nolongbibliography,
		citeautoscriptwocolumn,
]{revtex4-2}
\usepackage{graphicx} 
\usepackage{dcolumn} 
\usepackage{colordvi}
\usepackage{color}
\usepackage{epstopdf}
\usepackage{amssymb}
\usepackage{url}
\graphicspath{{ps}}
\usepackage{hyperref}
\usepackage{tabularx}
\usepackage{verbatim}
\usepackage{multirow}
\usepackage{siunitx}
\usepackage{soul}
\usepackage{orcidlink}

\input belle2sym.tex

\begin{document}

\def\belletwo {\it {Belle II}}
\def\Xbar {{\kern 0.18em\overline{\kern -0.18em X}{}\xspace}}
\newcommand{\detailtexcount}[1]{%
 \immediate\write18{texcount -merge -sum -q #1.tex output.bbl > #1.wcdetail }%
 \verbatiminput{#1.wcdetail}%
}

\newcommand{\quickwordcount}[1]{%
 \immediate\write18{texcount -1 -sum -merge -q #1.tex output.bbl > #1-words.sum }%
 \input{#1-words.sum} words%
}

\newcommand{\quickcharcount}[1]{%
 \immediate\write18{texcount -1 -sum -merge -char -q #1.tex output.bbl > #1-chars.sum }%
 \input{#1-chars.sum} characters (not including spaces)%
}

\title{Measurement of branching fractions and direct \textbf{\textit{CP}} asymmetries for $\boldsymbol{B \to K\pi}$ and $\boldsymbol{B\to\pi\pi}$ decays at Belle II}

  \author{I.~Adachi\,\orcidlink{0000-0003-2287-0173}} 
  \author{L.~Aggarwal\,\orcidlink{0000-0002-0909-7537}} 
  \author{H.~Ahmed\,\orcidlink{0000-0003-3976-7498}} 
  \author{H.~Aihara\,\orcidlink{0000-0002-1907-5964}} 
  \author{N.~Akopov\,\orcidlink{0000-0002-4425-2096}} 
  \author{A.~Aloisio\,\orcidlink{0000-0002-3883-6693}} 
  \author{N.~Anh~Ky\,\orcidlink{0000-0003-0471-197X}} 
  \author{D.~M.~Asner\,\orcidlink{0000-0002-1586-5790}} 
  \author{H.~Atmacan\,\orcidlink{0000-0003-2435-501X}} 
  \author{T.~Aushev\,\orcidlink{0000-0002-6347-7055}} 
% ...
\collaboration{The Belle II Collaboration}

\date{January 05, 2024}

\begin{abstract}
We report measurements of the branching fractions and direct $\it{CP}$ asymmetries of the decays $\Bz \to \Kp \pim$, $\Bp \to \Kp \piz$, $\Bp \to \Kz \pip$, and $\Bz \to \Kz \piz$, and use these for testing the standard model through an isospin-based sum rule. In addition, we measure the branching fraction and direct $\it{CP}$ asymmetry of the decay $\Bp \to \pip\piz$ and the branching fraction of the decay $\Bz \to \pip\pim$. The data are collected with the Belle II detector from $\epem$ collisions at the $\FourS$ resonance produced by the SuperKEKB asymmetric-energy collider and contain $387\times 10^6$ bottom-antibottom meson pairs. 
Signal yields are determined in two-dimensional fits to background-discriminating variables, and range from 500 to 3900 decays, depending on the channel.
We obtain $-0.03 \pm 0.13 \pm 0.04$ for the sum rule, in agreement with the standard model expectation of zero and with a precision comparable to the best existing determinations.

\end{abstract}

\maketitle

{\renewcommand{\thefootnote}{\fnsymbol{footnote}}}
\setcounter{footnote}{0}

\section{Introduction}
In the standard model (SM), charmless hadronic $B$ meson decays feature non-negligible contributions from loop amplitudes. 
These amplitudes are sensitive to contributions from non-SM physics. 
However, nonfactorizable amplitudes make application of perturbation theory difficult, and thus theory predictions tend to have large uncertainties.
Dynamical symmetries such as isospin symmetry can be exploited to construct sum rules, \emph{i.e.}, linear combinations of branching fractions and {\it CP} asymmetries, which reduce the impact of theoretical uncertainties~\cite{Gronau_2005}.
For the complete set of $B \to K\pi$ decays, $\Bz \to \Kp\pim$, $\Bp \to K^0\pip$, $\Bp \to \Kp\piz$, and $\Bz \to K^0\piz$~\footnote{Charge conjugation is implied throughout this paper unless stated otherwise.},
the sum rule parameter
\begin{equation}
\label{equ:isorule}
\begin{split}
  I_{K\pi} 
  &= \mathcal{A}_{{\it CP}}^{K^+\pi^-} 
  + \mathcal{A}_{{\it CP}}^{K^0\pi^+} \frac{\mathcal{B}_{K^0\pi^+}}{\mathcal{B}_{K^+\pi^-}}\frac{\mathcal{\tau}_{B^0}}{\mathcal{\tau}_{B^+}} \\
  &- 2 {\mathcal{A}_{{\it CP}}^{K^+\pi^0}} \frac{{\mathcal{B}_{K^+\pi^0}}}{\mathcal{B}_{K^+\pi^-}} \frac{\mathcal{\tau}_{B^0}}{\mathcal{\tau}_{B^+}} 
  - 2 \mathcal{A}_{{\it CP}}^{K^0\pi^0}\frac{\mathcal{B}_{K^0\pi^0}}{\mathcal{B}_{K^+\pi^-}}
\end{split}
\end{equation}
 is predicted to be zero within 1\% in the SM~\cite{PhysRevD.59.113002, bell_2015, bell_2020}, offering a reliable and sensitive null test.
Here, $\mathcal{A}_{{\it CP}}^{K\pi}$ and $\mathcal{B}_{K\pi}$ (with $K$ and $\pi$ charged or neutral)
 are the direct $\it{CP}$ asymmetry and the $\it{CP}$-averaged branching fraction of a $B \to K\pi$ decay, and $\mathcal{\tau}_{B^0}$ and $\mathcal{\tau}_{B^+}$ are the lifetimes of the neutral and charged $B$ mesons. The time-integrated $\it{CP}$ asymmetry is defined as
 \begin{equation}
 \label{eq:ACP}
   \mathcal{A}_{{\it CP}}^{X} = \frac{\Gamma(\Bbar \to \Xbar) - \Gamma(B \to X)}{\Gamma(\Bbar \to \Xbar) + \Gamma(B \to X)},
 \end{equation}
 where $\Gamma$ is the decay width to a specific final state $X$. 
 The asymmetry $\mathcal{A}_{{\it CP}}^{X}$ corresponds to the direct $\it{CP}$ asymmetry for charged $B$ mesons, and also for neutral $B$ mesons in the limit of no $\it{CP}$ violation in flavor oscillations, which is an excellent approximation for $\Bz$--$\Bzb$ mixing. 
The current value of the sum rule parameter is 
$I_{K\pi} = -0.13 \pm 0.11$, 
based on averages of measurements by the Belle, \babar, and LHCb collaborations~\cite{HFLAV}.
The sensitivity of the sum rule test is currently limited by the uncertainty on $\mathcal{A}_{{\it CP}}^{K^0\piz}$~\cite{PDG_2022}.

Isospin symmetry is also used in $B\to \pi\pi$ decays to determine the angle $\phi_2\equiv \mathrm{arg}[-{V_{td}V_{tb}^{*}}/({V_{ud}V_{ub}^{*}})]$, where $V_{ij}$ are the elements of the Cabibbo-Kobayashi-Maskawa (CKM) quark-mixing matrix~\cite{PhysRevLett.10.531, Kobayashi_1973}. The precision on our knowledge of $\phi_2$, also known as $\alpha$, is a limiting factor for testing the unitarity of the CKM matrix.
The measurement of the time-dependent asymmetry between rates of $\Bz$ and $\Bzb$ decays into $\pip \pim$ final states provides access to the angle $\phi_2$ up to an unknown shift.  The latter is constrained using isospin symmetry relations that require precise measurements of the branching fractions and $\it{CP}$ asymmetries of the decays $\Bz \to \pip \pim$, $\Bp \to \pip \piz$, and $\Bz \to \piz \piz$~\cite{PhysRevLett.65.3381, Charles_2017}.

In this paper, we report measurements of the branching fractions of $\Bz \to \Kp \pim$, $\Bp \to \Kp \piz$, $\Bp \to \Kz \pip$, $\Bz \to \Kz \piz$, $\Bz \to \pip \pim$, and $\Bp \to \pip \piz$ decays, and the direct {\it CP} asymmetries for all modes except ${\Bz \to \pip \pim}$. 
We use data corresponding to an integrated luminosity of $(362 \pm 2)\invfb$ collected with the Belle~II detector in asymmetric-energy electron-positron $(\epem)$ collisions at the $\FourS$ resonance provided by the SuperKEKB collider \cite{AKAI2018188}.
The data contains $(387\pm6)\times 10^6$ $\BB$ meson pairs.
This is the first measurement of the full set of isospin-related $B \to K\pi$ decays at {Belle~II} and provides a test of the $I_{K\pi}$ sum rule with precision competitive with the best previous determinations from single experiments.

The analyses of all decay modes follow a similar strategy. 
We develop multivariate algorithms to suppress the major source of background, which is continuum $\epem \to \qqbar$ processes, where $q$ indicates a $u$, $d$, $s$, or $c$ quark. 
We measure signal yields with likelihood fits to two signal-discriminating variables. 
We determine the {\it CP} asymmetries from charge-dependent signal-yield asymmetries for the flavor-specific decay modes. 
In $\Bz\to\Kz\piz$ decays, algorithms infer the production flavor of the signal $B$ meson using information from the other charged particles in the event. 
We use simulated events  to study the sample composition, to determine signal efficiencies, and to develop the fit model. 
We correct fit models and efficiencies for differences between data and simulation using several control channels, {\it e.g.}, $\Bp\to \Dzb(\to\Kp\pim) \pip$ decays. 
We develop and finalize all analysis choices and procedures using simulated and control-sample data before examining the signal data sample. We obtain a measurement of $I_{K\pi}$, where for $\mathcal{B}_{\Kz\piz}$ and $\mathcal{A}_{\it CP}^{\Kz\piz}$ we combine the results obtained in this work with the Belle II results on the time-dependent yield asymmetries~\cite{sagr:kspi0} to enhance sensitivity. 

\section{Detector and Samples}

The Belle II detector consists of subsystems arranged cylindrically around the interaction region~\cite{Belle-II:2010dht}. Belle II uses cylindrical coordinates in which the $z$ axis is approximately collinear with the electron beam.
Charged-particle trajectories (tracks) are reconstructed by a two-layer silicon-pixel detector~(PXD) surrounded by a four-layer double-sided silicon-strip
detector and a central 56-layer drift chamber.
The latter two detectors also measure the ionization energy loss. 
For data used in this work, one sixth of the second PXD layer was installed.
A quartz-based Cherenkov counter measures both the direction and time-of-propagation of photons
and identifies charged hadrons in the central region, and an aerogel-based ring-imaging Cherenkov counter identifies charged hadrons in the forward region.
An electromagnetic calorimeter made of CsI(Tl) crystals measures photon and electron energies and directions. 
The above subdetectors are immersed in a \SI{1.5}{T} axial magnetic field provided by a superconducting solenoid.
A subdetector dedicated to identifying muons and \KL mesons is installed outside of the solenoid. 

Large samples of simulated data are produced to model the physics processes resulting from $\epem$ collisions and to propagate the final state particles through a detailed simulation of the detector.
We use a series of software packages to produce the simulated data: \texttt{KKMC} to generate
continuum background~\cite{Jadach:1999vf}, \texttt{PYTHIA8} to simulate hadronization ~\cite{Sjostrand:2014zea}, \texttt{EVTGEN} to simulate decays~\cite{Lange:2001uf}, \texttt{PHOTOS} to simulate final state radiation \cite{BARBERIO1991115}, and \texttt{GEANT4} to model the detector response~\cite{GEANT4:2002zbu}. 
Our simulation includes overlays of beam-induced backgrounds~\cite{Liptak_2022} that correspond to the conditions of the data we use.
Collision and simulated data are processed with the Belle II software~\cite{Kuhr:2018lps, Basf2-zenodo}.

\section{Sample selection}

All events are required to pass loose online selection criteria based on the total energy and charged-particle multiplicity in the event. 
In the offline analysis, tracks are combined with particle-identification information (PID) to select charged pion and kaon candidates. 
These are combined in kinematic fits to reconstruct the desired signal decay chains.
All tracks must have a polar angle within the drift-chamber acceptance ${[17^{\circ}, 150^{\circ}]}$ and be associated with 20 or more measurement points (hits). Tracks not used to form a \KS candidate are required to have a distance of closest approach to the interaction point of less than 2.0~cm along the $z$ axis and less than 0.5~cm in the transverse plane.
For the $\Bz \to h^+ \pim$ and $\Bp \to h^+ \piz$ decays, where $h$ indicates a kaon or a pion, we separate the data into independent pion- and kaon-enriched samples based on PID.  We use the ratio $\mathcal{L}_{\pi}/(\mathcal{L}_{\pi}+ \mathcal{L}_{K})$, where the likelihood $\mathcal{L}_i$ for a pion or kaon hypothesis combines PID information from all subdetectors except the PXD. We discard events in which both tracks are identified as kaons.
In the $\Bz \to h^+ \pim$ modes, the PID requirement correctly identifies 90\% of pions and 84\% of kaons.
In the $\Bp \to h^+ \piz$ modes, 86\% of both pions and kaons are correctly identified.
Due to different backgrounds in the two samples, the PID requirement varies between the $\Bz \to h^+ \pim$ and $\Bp \to h^+ \piz$ modes, which results in different identification efficiencies. 

Neutral pion candidates are reconstructed from pairs of photon candidates, which are detected in the electromagnetic calorimeter as sets of adjacent crystals with energy deposits (clusters) that are not associated to a track. 
Clusters are required to involve more than one crystal, to have an energy deposit greater than $30\mev$, and a time within $200\ns$ of the estimated event time.
We apply a requirement on the output of a boosted decision-tree (BDT) as described in Ref.~\cite{francis:pi0pi0} to suppress misreconstructed photons.
For $\piz$ candidates, the angle between the photons is required to be smaller than 0.4 radian, and the diphoton mass must satisfy ${115 < m({\gamma\gamma}) < 150\mevcc}$.
This range corresponds to approximately $2.5\sigma$ around the known $\piz$ mass~\cite{PDG_2022}, where $\sigma$ is the $m({\gamma\gamma})$ resolution.

Neutral kaon candidates are reconstructed via their $\KS\to \pip\pim$ decays by combining pairs of oppositely charged particles (assumed to be pions) with the dipion mass $480 < m(\pi\pi)<510\mevcc$. 
This range corresponds to more than $\pm 3\sigma$, where $\sigma$ is the resolution around the known $\KS$ mass~\cite{PDG_2022}.
We use two BDTs to select \KS candidates. The first suppresses combinatorial background and uses 15 input variables that include the kinematic information of the \KS and associated $\pip\pim$ candidates. 
The most discriminating variables are the flight length of the \KS normalized by its uncertainty, and the angle between the direction of the \KS momentum and the vector connecting the interaction point and the vertex position.
The second BDT reduces background from $\Lambda$ decays. It employs five input variables that capture kinematic information for the associated $\pi$ candidates, PID information, and the invariant masses obtained when either one of the tracks is assumed to be a proton. The latter is the most discriminating input of the second BDT. 
The requirements on $\piz$ and $\KS$ candidates are chosen to maximize the ratio ${\rm S}/\sqrt{{\rm S+B}}$, where S and B are the signal and continuum background event yields obtained in simulated samples.

Signal $B$ decay candidates are selected using  
the beam-constrained mass $M_{\text{bc}} \equiv \sqrt{(E^{*}_{\text{beam}}/c^2)^2 - (|\boldsymbol{p}^{*}_{B}|/c)^2}$, and the energy difference $\Delta E \equiv E^{*}_{B} - E^{*}_{\text{beam}}$. The beam energy $E^{*}_{\text{beam}}$ and $B$ meson momentum $\boldsymbol{p}^{*}_{B}$ and energy $E^{*}_{B}$ are calculated in the center-of-mass frame, as are all starred quantities henceforth. Correctly reconstructed signal events peak at the $B$ meson mass~\cite{PDG_2022} in $M_{\text{bc}}$ and at zero in $\Delta E$. 
Misreconstructed $B$ decays in which incorrect mass hypotheses are assumed for the final state particles peak at the $B$ meson mass in $M_{\text{bc}}$ but are shifted in $\Delta E$ by typically 30--50\mev.

In order to improve the $M_{\text{bc}}$ resolution for decays containing a $\piz$, in the calculation of $M_{\text{bc}}$ we replace the measured $\piz$ momentum with ${\boldsymbol{p}^{*\prime}_{\piz} = \sqrt{(E^{*}_{\text{beam}} - E^{*}_{h})^2/c^2 - m^2_{\piz}c^2} \times \frac{\boldsymbol{p}^{*}_{\piz}}{|\boldsymbol{p}^{*}_{\piz}|}}$,
where $E^{*}_{h}$ is the energy of the charged kaon, charged pion, or neutral kaon from the $B$ candidate decay, $\boldsymbol{p}^{*}_{\piz}$ is the measured $\piz$ momentum, and $m_{\piz}$ is the known $\piz$ mass~\cite{PDG_2022}.
Signal simulations show that the $M_{\text{bc}}$ resolution is improved by a factor of about $1.2$.
This substitution also reduces correlations between $M_{\text{bc}}$ and $\Delta E$ for two-body final states with a $\piz$~\cite{PhysRevD.87.031103}.

Candidate $B$ decays are retained if they satisfy $5.272 < M_{\text{bc}} < 5.288\gevcc$ and $-0.3 < \Delta E < 0.3\gev$. 
For the $\Bz \to h^+ \pim$ decay modes, which exhibit a better momentum resolution than the other modes, a more restrictive requirement $-0.1 < \Delta E < 0.2\gev$ is applied to suppress background from partially reconstructed multibody decays populating the region $\Delta E <-0.1\gev$.

After these selections, a large contribution from continuum background remains.
To reject this background we train BDT continuum suppression (CS) discriminators, separately for the $\Bz \to h^+ \pim$, $\Bz \to \KS \piz$, and $\Bp \to h^+ \piz$ modes.
The variables used in the training are related to the event shape (modified Fox-Wolfram moments~\cite{fw, ABE2001151, PhysRevLett.87.101801}, CLEO cones~\cite{CLEO:1995rok}, sphericity-related quantities~\cite{sphericity}, and thrust-related quantities~\cite{thrust_related}), flavor-tagger output~\cite{Belle-II:2021zvj} (except for $\Bz\to\KS\piz$), results of the $B$ meson decay-vertex fit, and $M_{\text{bc}}$.
Variables whose correlation with $\Delta E$ exceeds 5\% are excluded from the inputs.
We apply loose requirements on the BDT output to remove 90\%--99\% of continuum background while retaining 78\%--96\% of signal decays. 
The BDT output is then used as a fitting observable.

After all selection criteria are applied, the fraction of events with multiple candidates is less than 1\% for all channels. We retain all candidates in the rest of the analysis. 

\section{Determination of the branching fractions and direct {\it \textbf{CP}} asymmetries}
\label{sec:fit}
To determine branching fractions and {\it CP} asymmetries of the decays, we fit the unbinned two-dimensional distribution of $\Delta E$ and $C'$. The latter is the output of the CS BDT transformed using the probability integral transformation \cite{TMVA} such that the signal has a uniform distribution from zero to one, while the continuum is found empirically to be well-modeled by an exponential distribution. 

We perform four extended maximum-likelihood fits: two simultaneous fits to the pion- and kaon-enriched samples, one for the ${\Bz \to h^+ \pim}$ decays and the other for $\Bp \to h^+ \piz$ decays; 
a fit to the sample of $\Bp \to \KS \pip$ decays; and a fit to the sample of $\Bz \to \KS \piz$ decays. For the $\Bz \to K^+ \pim$ and $\Bz \to \pi^+ \pim$ decays, fitting simultaneously to the pion- and kaon-enriched samples properly accounts for candidates in the complementary sample in which one final state particle is misidentified; we refer to such candidates as feed across.
For the same reason, we perform a simultaneous fit to the $\Bp \to \Kp \piz$ and $\Bp \to \pip \piz$ subsamples.
The feed-across reconstruction efficiencies are fixed to the values determined from large simulation samples.

\begin{table*}[tb]
  \caption{Signal yields, feed-across yields, signal reconstruction efficiencies, feed-across reconstruction efficiencies, branching fractions, and direct $\it{CP}$ asymmetries. The signal and feed-across reconstruction efficiencies are corrected for differences between data and simulation in the CS efficiency, $\KS$ reconstruction efficiency, and PID efficiency. They are also multiplied by the subdecay branching fractions, which are 0.5 times the $\KS \to\pip\pim$ branching fraction for decays with a $\Kz$, and the $\piz\to\gamma\gamma$ branching fraction for those with a $\piz$.
  The first (or sole) contribution to uncertainties denotes the statistical component, the second denotes the systematic component. 
  }
  \label{tab:fit_result}
  \begin{tabular}{lcl rcl rcl c c rcl cl rcl cl}
  \hline
  \hline
  \multicolumn{3}{l}{\multirow{2}{*}{Decay}} & \multicolumn{3}{c}{Signal} & \multicolumn{3}{c}{Feed-across} & Signal & Feed-across & \multicolumn{5}{c}{\multirow{2}{*}{$\mathcal{B}$ $[10^{-6}]$}} & \multicolumn{5}{c}{\multirow{2}{*}{$\mathcal{A}_{\it{CP}}$}} \\
   & & & \multicolumn{3}{c}{yield} & \multicolumn{3}{c}{yield} & $\epsilon$ [\%] & $\epsilon$ [\%] & & \\
  \hline
  $\Bz$ & $\to$ &$\Kp \pim$ & 3868 & $\pm$ & 71 & 880 & $\pm$ & 16 & 49.91 & 11.37 & 20.67 & $\pm$ & 0.37 & $\pm$ & 0.62 & $-0.072$ & $\pm$ & 0.019 & $\pm$ & 0.007 \\
  $\Bz$ &$\to$ &$\pip\pim$ & 1187 & $\pm$ & 43 & 327 & $\pm$ & \phantom{1}8 & 54.31 & 14.94 & 5.83 & $\pm$ & 0.22 & $\pm$ & 0.17 & \multicolumn{5}{c}{$\cdot\cdot\cdot$} \\
  $\Bp$&$ \to$&$ \Kp \piz$ & 2052 & $\pm$ & 57 & 359 & $\pm$ & 10 & 36.91 & \phantom{1}6.46 & 13.93 & $\pm$ & 0.38 & $\pm$ & 0.71 & 0.013 & $\pm$ & 0.027 & $\pm$ & 0.005 \\
  $\Bp$&$ \to$&$ \pip \piz$ & 785 & $\pm$ & 44 & 136 & $\pm$ & \phantom{1}8 & 37.60 & \phantom{1}6.50 & 5.10 & $\pm$ & 0.29 & $\pm$ & 0.27 & $-0.081$ & $\pm$ & 0.054 & $\pm$ & 0.008 \\
  $\Bp$&$ \to$&$ \Kz \pip$ & 1547 & $\pm$ & 45 & &  & & 15.89 &  & 24.37 & $\pm$& 0.71 & $\pm$ & 0.86 & 0.046 & $\pm$ & 0.029 & $\pm$ & 0.007 \\
  $\Bz$&$ \to$&$ \Kz \piz$ & \multirow{2}{*}{502} & \multirow{2}{*}{$\pm$} & \multirow{2}{*}{32} & \multirow{2}{*}{} & \multirow{2}{*}{} & \multirow{2}{*}{} & 
  \multirow{2}{*}{12.38} & \multirow{2}{*}{} & \multirow{2}{*}{10.40} & \multirow{2}{*}{$\pm$} & \multirow{2}{*}{0.66} & \multirow{2}{*}{$\pm$} & \multirow{2}{*}{0.60} & \multirow{2}{*}{$-0.06$} & \multirow{2}{*}{$\pm$} & \multirow{2}{*}{0.15} & \multirow{2}{*}{$\pm$} & \multirow{2}{*}{0.04} \\ 
   \multicolumn{3}{l}{(this analysis)} & & & & & & & & \\

  $\Bz$ & $\to$ & $\Kz \piz$ &\multirow{2}{*}{415} & \multirow{2}{*}{$\pm$} & \multirow{2}{*}{26} & &\multirow{2}{*}{} & &\multirow{2}{*}{\phantom{5}9.87} & \multirow{2}{*}{} & \multirow{2}{*}{11.15} & \multirow{2}{*}{$\pm$} & \multirow{2}{*}{0.68} & \multirow{2}{*}{$\pm$} & \multirow{2}{*}{0.62} & \multirow{2}{*}{$0.04$} & \multirow{2}{*}{$\pm$} & \multirow{2}{*}{0.15} & \multirow{2}{*}{$\pm$} & \multirow{2}{*}{0.05} \\
   \multicolumn{3}{l}{(time-dependent analysis~\cite{sagr:kspi0})} & & & & & & \\
   $\Bz$ & $\to$ & $\Kz \piz$ & & \multirow{2}{*}{} & & &\multirow{2}{*}{} & & \multirow{2}{*}{} & \multirow{2}{*}{} & \multirow{2}{*}{10.73} & \multirow{2}{*}{$\pm$} & \multirow{2}{*}{0.63} & \multirow{2}{*}{$\pm$} & \multirow{2}{*}{0.62} & \multirow{2}{*}{$-0.01$} & \multirow{2}{*}{$\pm$} & \multirow{2}{*}{0.12} & \multirow{2}{*}{$\pm$} & \multirow{2}{*}{0.04} \\
   \multicolumn{3}{l}{(combination with Ref.~\cite{sagr:kspi0})} & & & & & & \\
  \hline
  \hline
  \end{tabular}
\end{table*}

To measure $\mathcal{A}_{{\it CP}}$, we divide each sample into two subsamples according to the signal candidate flavor, and fit the subsamples simultaneously. To determine the flavor, the charge of the reconstructed $B$ candidate is used for $B^+$ decays, and that of the kaon for the self-tagging ${\Bz \to \Kp \pim}$ decay. 
For the ${\Bz \to \KS \piz}$ decay, the production flavor of the signal $B$ is determined by reconstructing the accompanying (tag-side) $B$ meson in each event using a category-based algorithm~\cite{Belle-II:2021zvj}. 
The tagging information is characterized by two parameters: the flavor of the tag-side $B$ 
and an event-by-event dilution factor $r \equiv 1 - 2\omega$, where $\omega$ is the probability that an event is mistagged. 
The dilution factor ranges from zero, for no flavor distinction between $\Bz$ and $\Bzb$, to one for an unambiguous flavor assignment. 
The asymmetry $\mathcal{A}_{{\it CP}}$ is diluted by a factor $r$ due to incorrect tagging.
We divide the sample into seven intervals in $r$ and fit the seven subsamples simultaneously.
Since the $r$ distribution differs between signal and continuum, we gain additional statistical sensitivity in the fit. In addition, we enhance sensitivity to the signal asymmetry by using the average dilution in each $r$ interval instead of the average dilution of the entire sample.  
The observed asymmetry of ${\Bz \to \Kz \piz}$ is reduced by an additional factor of $(1-2\chi_d)$, where $\chi_d = 0.1858\pm 0.0011$ is the decay-time-integrated $\Bz$--$\Bzb$ mixing probability~\cite{PDG_2022}.

For each channel, we consider three sample components: signal, continuum background, and background from other $B$ decays (referred to as $\BB$ background). The continuum background is dominant in all samples. 
The \BB background features partially reconstructed decays that contribute to the region below $\Delta E< -0.1\gev$. This background is reduced to a negligible level for $\Bz \to h^+ \pim$ as a result of the narrower $\Delta E$ window used in the selection, but must be considered for other decay modes.
For $\Bz\to h^+\pim$ and $\Bp\to h^+\piz$ decays, there is an additional peaking component from feed across.
For ${\Bp \to \KS \pip}$, the $\Bp \to \KS \Kp$ decay is included as an additional component in the fit that originates from kaon-to-pion misidentification and peaks close to the signal.

The joint probability density function (PDF) of $\Delta E$ and $C'$ is assumed to factorize into the product of two independent one-dimensional PDFs. This assumption is supported by simulation studies for each fit. For the ${\Bz \to \KS \piz}$ decay, we check that the signal and background shapes are the same for each $r$ interval. 
Fit shapes are determined empirically from large simulation samples as follows. 
We model the signal shapes in $\Delta E$ with crystal Ball \cite{Oreglia:1980cs} and Gaussian functions, and use linear functions for $C'$. 
We use similar models for the feed-across and peaking background components. 
We model the continuum with first- or second-order polynomials in $\Delta E$ and exponential functions in $C'$. 
For the $\BB$ background, we use either nonanalytic PDFs defined using kernel estimation~\cite{Cranmer_2001} or Gaussian functions for $\Delta E$, and linear functions for $C'$.

For all modes, shape parameters for signal and peaking backgrounds are fixed from fits to simulated samples. 
For the $\Delta E$ signal and peaking-background shape, two parameters are used to correct for data-simulation differences: a shift and a smearing factor for the Gaussian part of the shapes. 
These parameters are determined from large control samples, namely $\Bp \to \Dzb (\to \Kp \pim) \pip$ for $\Bz\to h^+\pim$, $\Bz \to \Dzb (\to \Kp \pim) \piz$ for $\Bp\to h^+\piz$ and $\Bz\to\KS\piz$, and $\Bp \to \Dzb (\to \KS \pip \pim) \pip$ as well as $\Bz \to \Dm (\to \KS \pim) \pip$ for $\Bp\to\KS\pip$ decays.
For all modes, the continuum-background shape parameters are unconstrained.

\begin{table*}[htb!]
    \centering
    \caption{Intervals of dilution $r$ and associated flavor parameters.}
    \label{tab:flv_param_data}
    \begin{tabular}{rcl c rcl rcl rcl}
    \hline
    \hline
    \multicolumn{3}{c}{$r$ interval} &&  \multicolumn{3}{c}{$w_{r}$}  & \multicolumn{3}{c}{$\Delta w_{r}$} &  \multicolumn{3}{c}{$\Delta\epsilon_{\text{tag},r}$} \\
    \hline
    0.000&$-$&0.100  &&  ~~~0.4804&$\pm$&0.0054  &  ~~~$-0.007$&$\pm$&0.011 & ~~~$-0.024$&$\pm$&0.012\\
    0.100 &$-$& 0.250  && 0.4240&$\pm$&0.0054  &  0.038&$\pm$&0.011 & 0.014&$\pm$&0.012 \\
    0.250 &$-$& 0.450  &&  0.3410&$\pm$&0.0051  &  $-0.019$&$\pm$&0.010& $-0.011$&$\pm$&0.012 \\
    0.450 &$-$& 0.600  &&  0.2362&$\pm$&0.0053  &  $-0.007$&$\pm$&0.011 &  0.009&$\pm$&0.013 \\
    0.600 &$-$& 0.725  &&  0.1675&$\pm$&0.0053  &  0.020&$\pm$&0.011 & 0.037&$\pm$&0.013   \\
    0.725 &$-$& 0.875  &&  0.1073&$\pm$&0.0048  &  0.000&$\pm$&0.010 & $-0.020$&$\pm$&0.013   \\
    0.875 &$-$& 1.000  &&  0.0274&$\pm$&0.0030  &  0.002&$\pm$&0.006 & $-0.012$&$\pm$&0.011 \\
    \hline\hline
    \end{tabular}
\end{table*}

In the fits, the signal yield to a final state $X$ is expressed as a function of the signal branching fraction $\mathcal{B}_X$ and the {\it CP} asymmetry $\mathcal{A}_{\it CP}^X$ as
\begin{equation}
\label{eq:signal_yields}
  N_X^{q} = 2\, \mathcal{N}\,  f^{+-/00} \, \epsilon_X \, \mathcal{B}_X\, \frac{ \Delta^q - \, \mathcal{D}^q\,\mathcal{A}_{C\!P}^X }{2},
\end{equation}
where $q = +1$ for a $B$ meson and $-1$ for a $\Bbar$ meson; $\mathcal{N}$ is the number of produced $\BB$ pairs, $387\times10^6$; $f^{+-/00}$ is the fraction of either $\Bp\Bm$ or $\BzBzb$ production at the $\Upsilon(4S)$, which is  $ 0.484 \pm 0.012$ for $\BzBzb$ and $0.516 \pm 0.012$ for $\Bp\Bm$~\cite{f+-/f00}; and $\epsilon_X$ is the charge-averaged signal efficiency determined from simulated signal samples. 
In the fits for $\Bz\to h^+\pim$ and $\Bp\to h^+\piz$ decays, the feed-across yields contribute to the determination of the signal branching fractions and asymmetries; for those yields, $\epsilon_{X}$ in Eq.~(\ref{eq:signal_yields}) is the feed-across efficiency determined from simulation.
The signal and feed-across efficiencies, listed in Table~\ref{tab:fit_result}, are corrected for data-simulation differences in the CS efficiency, $\KS$ reconstruction efficiency, and PID efficiency.
For decays with a $\KS$, a factor of 0.5 for the $K^{0} \to K^{0}_{S}$ probability is included in $\epsilon_X$, as well as the branching fraction of $\KS \to\pip\pim$~\cite{PDG_2022}; 
for decays with a $\piz$, the branching fraction of $\piz\to\gamma\gamma$ is included.
The factors $\Delta^q$ and $\mathcal{D}^q$ take the values one and $q$, respectively, for all cases except for the $\Bz \to \KS \piz$ decay, where they encode flavor-tagging asymmetries and dilution factors as follows:
\begin{equation}
\begin{split}
  \Delta^q &= 1 - q [\Delta \omega_r + \Delta \epsilon_{{\rm tag},r}(1-2 \omega_r)], \\
  \mathcal{D}^q &= (1 -2 \chi_d)[q(1-2\omega_r) + \Delta \epsilon_{{\rm tag},r}(1 - q \Delta \omega_r)],
\end{split}
\end{equation}
where $\omega_r$ is the average fraction of wrong-tag candidates in one of the seven intervals of dilution $r$, $\Delta \omega_r$ is the difference in the wrong-tag fraction between positive and negative tags, and
$\Delta \epsilon_{{\rm tag},r}$ is the asymmetry of the tagging efficiency. The fraction of signal events in each $r$ interval, along with $\omega_r$, $\Delta \omega_r$, and $\Delta \epsilon_{{\rm tag},r}$, are fixed to values determined from $B \to D^{(*)} h^+$ decays, following Ref.~\cite{*[{}]  [{. The parameters have been derived from the same methods used in that  analysis with the larger data set of this paper.}] lifemixing}.
These parameters are listed in Table~\ref{tab:flv_param_data}.

In all fits except for the ${\Bz \to \KS \piz}$ fit, separate background yields for each $B$ flavor are determined by the fit to account for possible background asymmetries. 
For the fit to the ${\Bz \to \KS \piz}$ sample, the backgrounds are assumed to be flavor symmetric.
Potential bias from this assumption is accounted for in the systematic uncertainty.
For the $\BB$ background, we use the same parameters associated with flavor-tagging as used for signal; for continuum, the fraction of candidates in each $r$ interval is fixed from simulation.

Corrections applied to the reconstruction efficiencies and direct {\it CP} asymmetries are discussed in the following section.

\section{Corrections}
\label{sec:corrections}
We correct the reconstruction efficiencies for data-simulation differences in the $\piz$ reconstruction, the $\KS$ reconstruction, the efficiency of the CS requirement, and the PID selection efficiency. 
The $\piz$ reconstruction efficiency is obtained by measuring the ratio of the yields of $\Dz \to \Km \pip \piz$ and $\Dz \to \Km \pip$ in data, using the known values of their branching fractions as input \cite{PDG_2022}. 
The corrections are derived as a function of the momentum and polar angle of the $\piz$ candidates to account for the different kinematic distributions of control and signal decays.  
The $\KS$ reconstruction efficiency is evaluated using $\Dstarp \to \Dz (\to \KS \pip \pim) \pip$ and $\Dstarp \to \Dz (\to \KS \piz) \pip$ decays, and the CS efficiency is determined using $\Bp \to \Dzb (\to \Kp \pim) \pip$, $\Bp \to \Dzb (\to \Kp \pim \piz) \pip$, and $\Bp \to \Dzb (\to \KS \piz) \pip$ decays.
Corrections for efficiencies and misidentification rates of the PID selections are obtained as functions of particle momentum and polar angle from abundant control samples of $\KS \to \pip \pim$ and $\Dstarp \to \Dz (\to \Km \pip) \pip$ decays. 
We measure the efficiencies in these channels for both data and simulated events, and subsequently scale the reconstruction efficiencies of our signal channels by the observed ratio of efficiencies.
Depending on the decay mode, the correction ranges from $-0.8\%$ to $-0.5\%$ for the $\piz$ reconstruction efficiency, $-6.2\%$ to $-4.8\%$ for the $\KS$ reconstruction efficiency, $-3.2\%$ to $-0.1\%$ for the CS efficiency, $-3.0\%$ to $-1.4\%$ for the PID efficiency, and $16\%$ to $20\%$ for the PID misidentification rate.

The raw asymmetries are corrected for differences in reconstruction efficiencies for particles and antiparticles that arise from the difference in their interaction probabilities~\cite{Collaboration:3332}.
We estimate the instrumental asymmetry for charged pions by measuring the charge asymmetry in a large sample of $\Dp \to \KS \pip$ decays assuming negligible contributions from \KS asymmetries and subtracting the known value of $\mathcal{A}_{\it{CP}}(\Dp \to \KS \pip)$ \cite{PDG_2022}. 
To obtain the instrumental asymmetry for charged kaons, 
we first determine the charge asymmetry in $\Dz \to \Km \pip$ decays; as this amplitude is Cabibbo-favored, direct {\it CP} violation in this channel is expected to be negligible. The instrumental asymmetry measured is due to both the charged kaon and the pion; we determine that due to the kaon by correcting the asymmetry by the pion asymmetry measured in $\Dp \to \KS \pip$ decays. 
The $D$ decays are selected to originate from $\epem\to\ccbar$ by requiring the momentum of the $D$ mesons in the center-of-mass frame to be greater than $2.5\gevc$.
Instrumental asymmetries depend on the kinematic distributions of final state particles; thus, charged particles in the control channels are required to have similar kinematic distributions as charged particles in the signal channels.
The instrumental asymmetry for the different signal decays ranges from 0.006--0.014.

\section{Results}
\begin{figure*}[htb]
  \centering
  \includegraphics[width=0.49\linewidth]{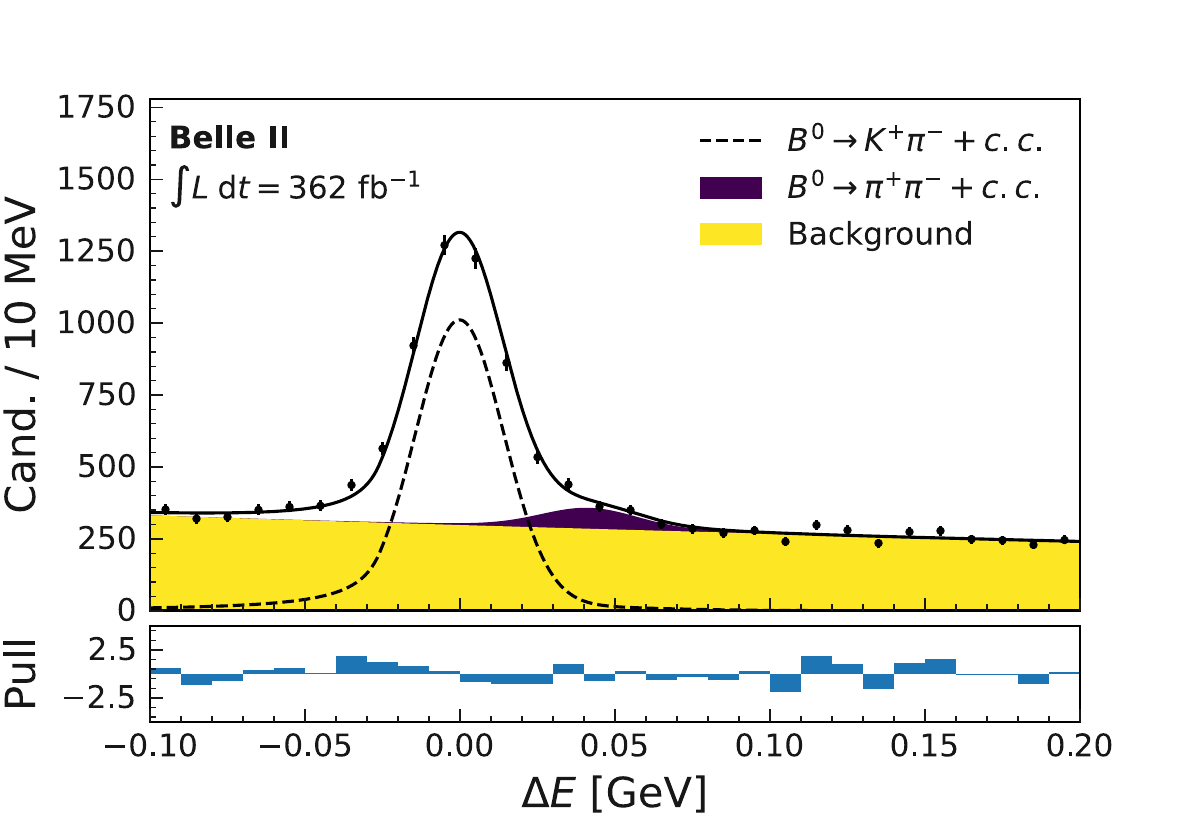}
  \includegraphics[width=0.49\linewidth]{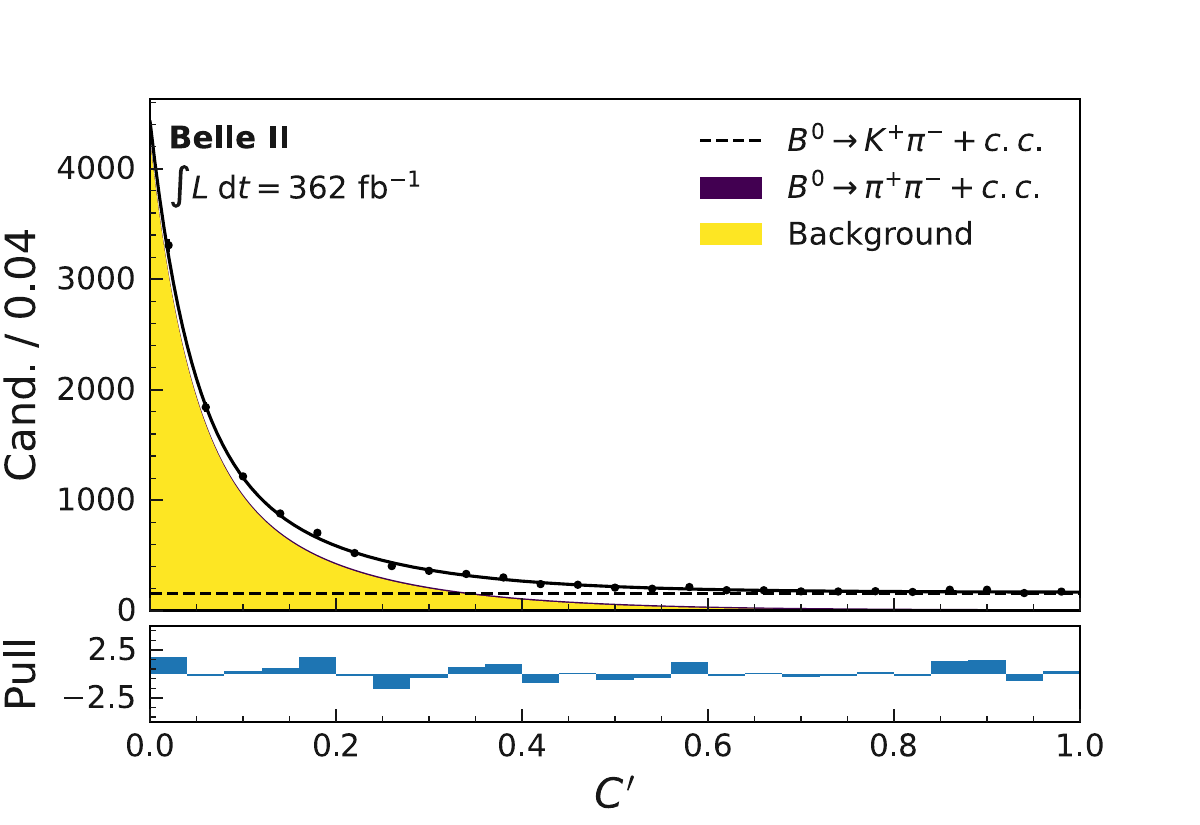}
  \includegraphics[width=0.49\linewidth]{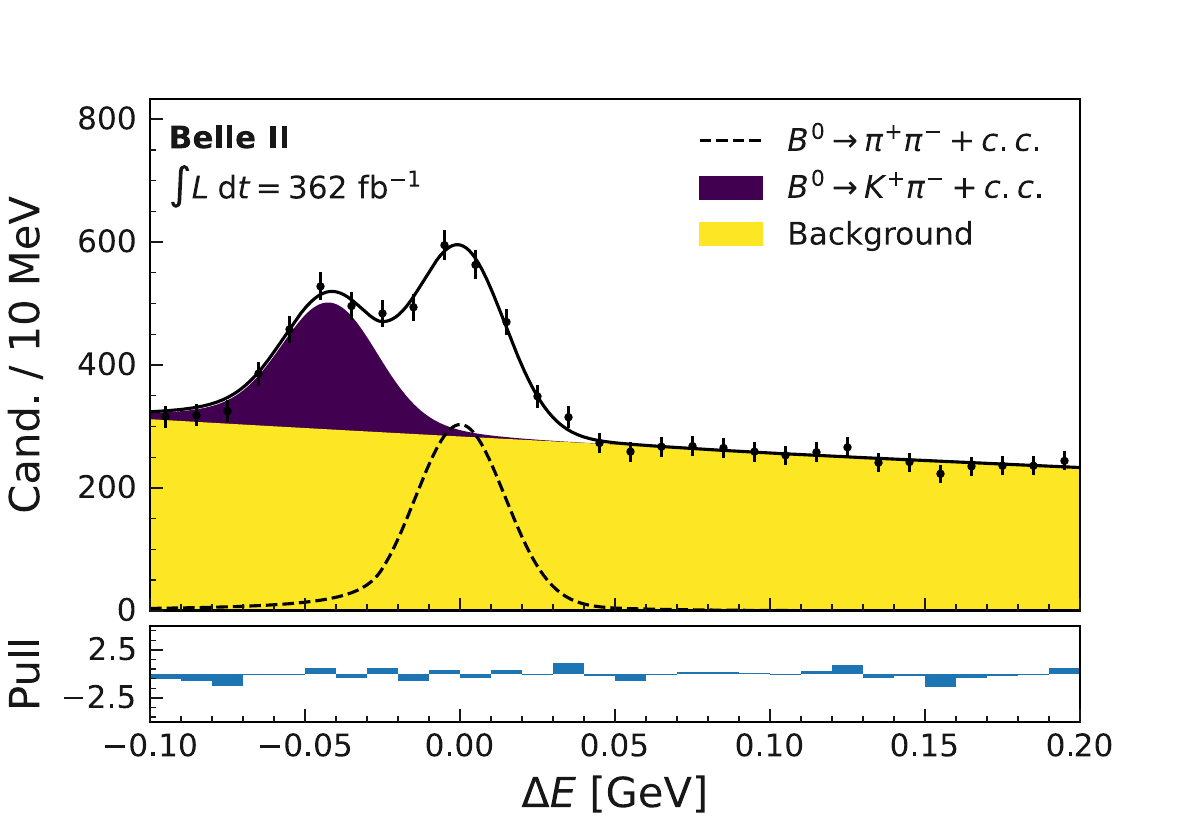}
  \includegraphics[width=0.49\linewidth]{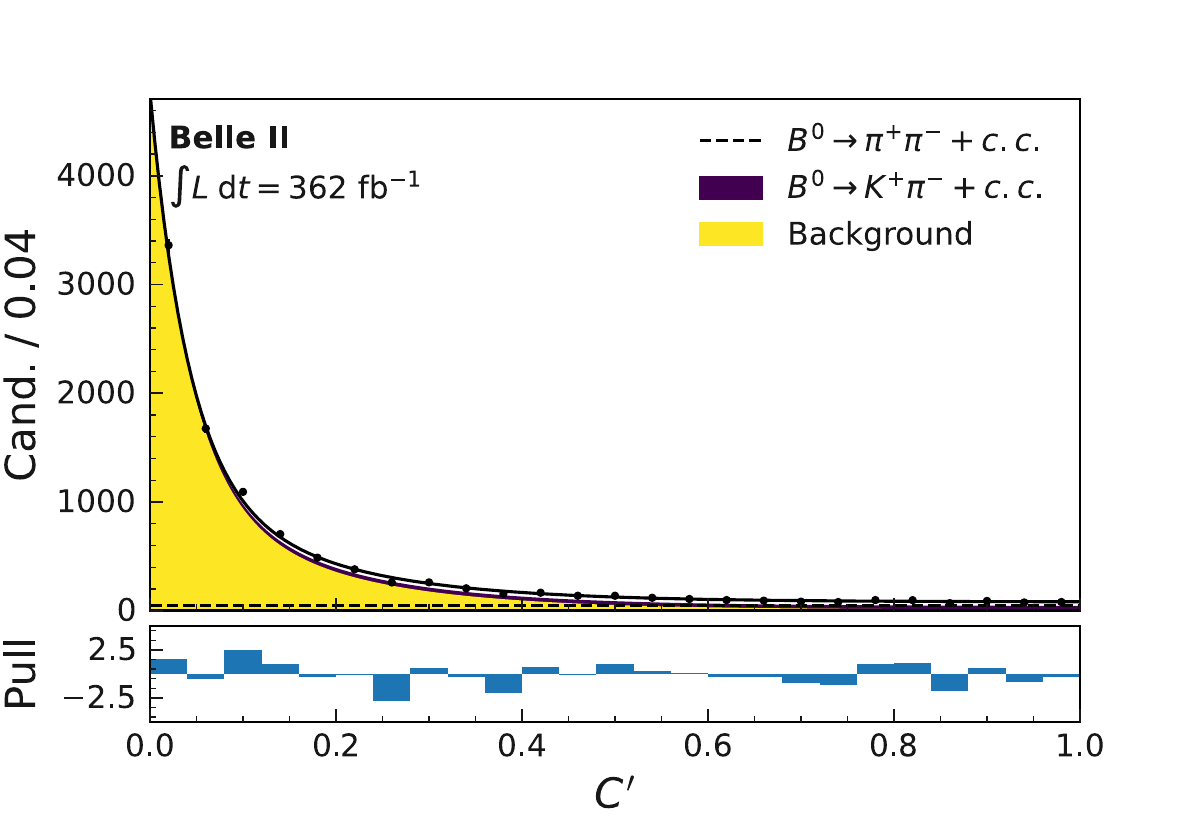}
  \caption{Distribution of (left) $\Delta E$ and (right) transformed CS output $C'$ for the (top) kaon- and (bottom) pion-enriched samples of $\Bz \to h^+ \pim$ candidates with fit results overlaid. Points with error bars are data, the total fit is shown as a solid black curve, the signal as a black dashed curve, the feed across as a purple shaded area, and background as a yellow shaded area. Differences between observed data and total fit results normalized by fit uncertainties (pulls) are also shown.
  }
  \label{fig:fit_result}
\end{figure*}

\begin{figure*}[htb]
  \centering
  \includegraphics[width=0.49\linewidth]{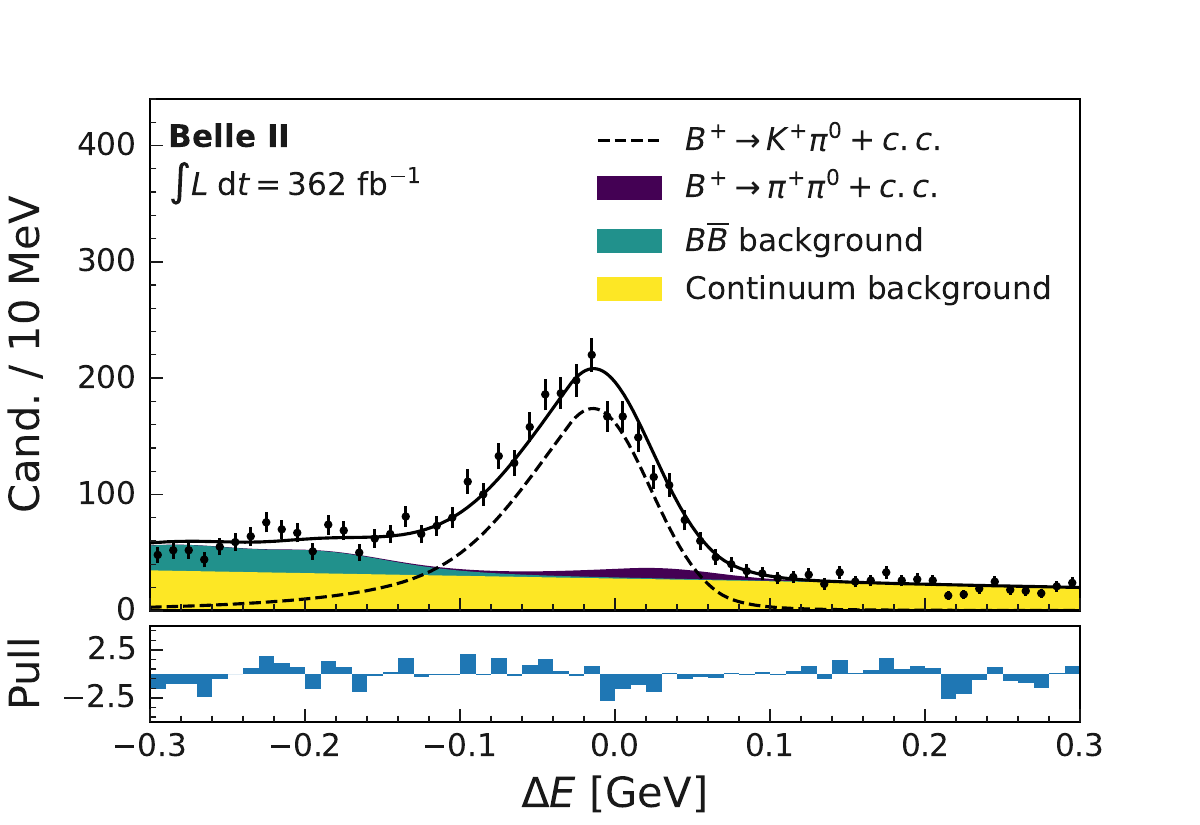}
  \includegraphics[width=0.49\linewidth]{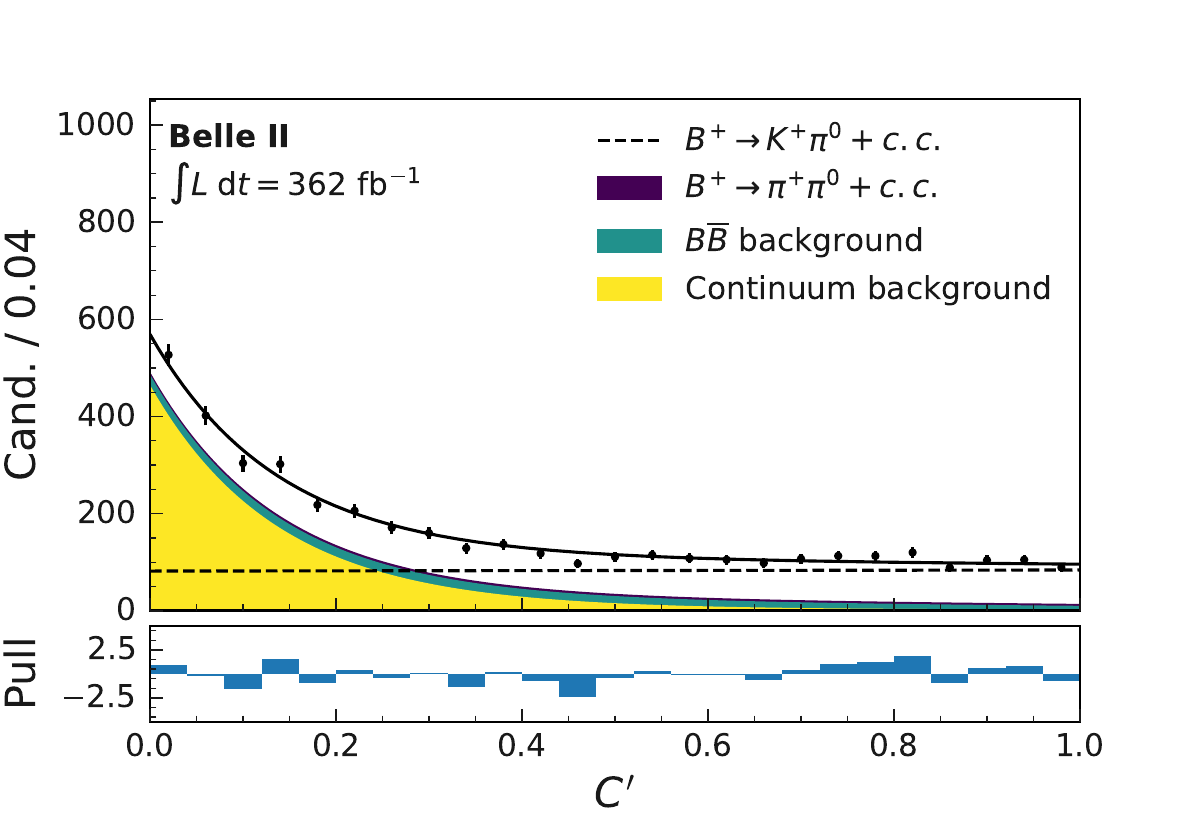}
  \includegraphics[width=0.49\linewidth]{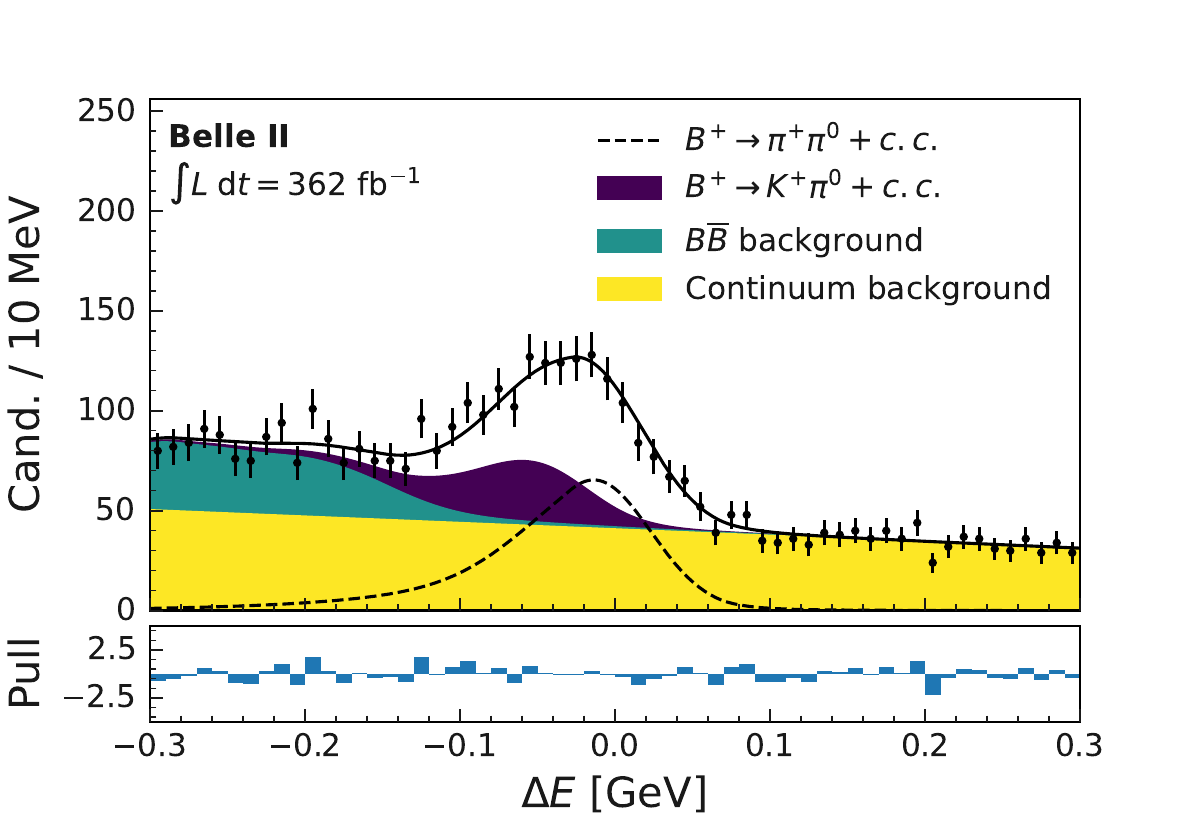}
  \includegraphics[width=0.49\linewidth]{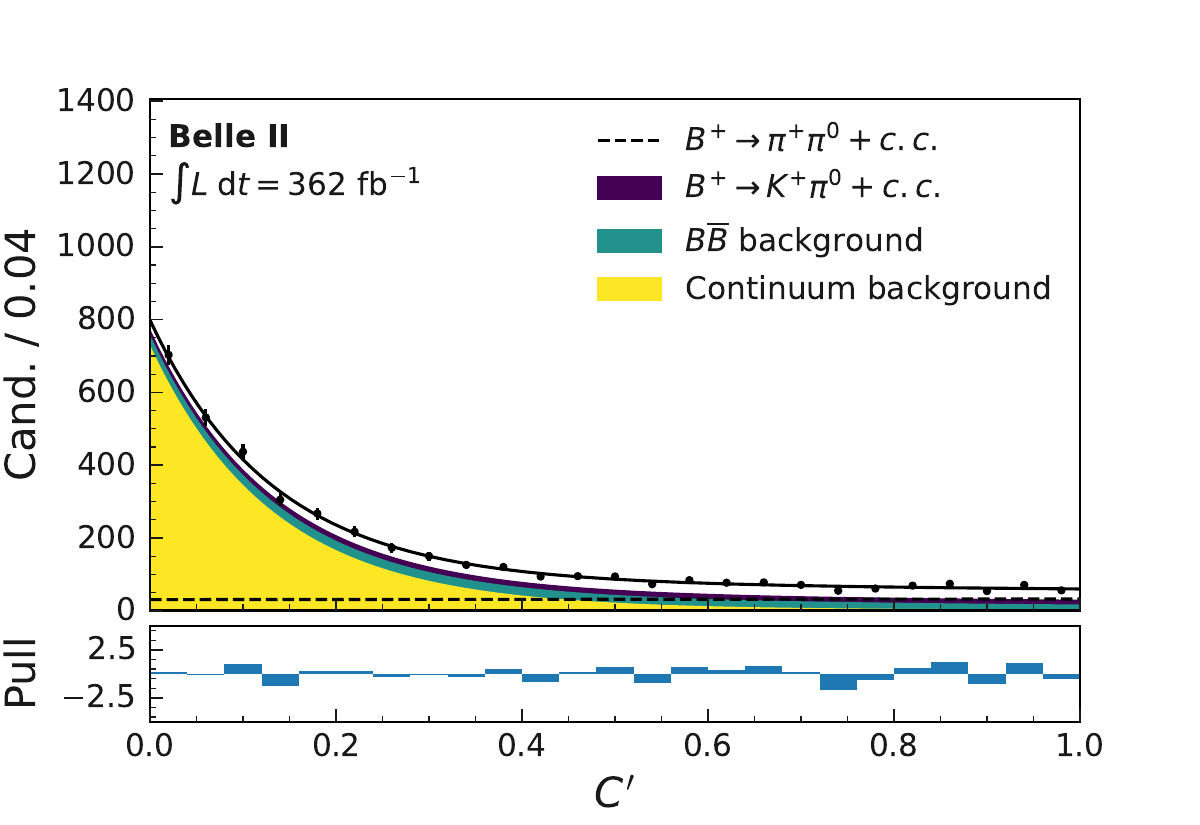}
  \caption{Distribution of (left) $\Delta E$ and (right) transformed CS output $C'$ for the (top) kaon- and (bottom) pion-enriched samples of $\Bz \to h^+ \piz$ candidates with fit results overlaid.  Points with error bars are data, the total fit is shown as a solid black curve, the signal as a black dashed curve, the feed across as a purple shaded area, the $\BB$ background as a green shaded area, and the continuum background as a yellow shaded area. Fit pulls are also shown.
  }
  \label{fig:fit_result_2}
\end{figure*}

\begin{figure*}[htb]
  \centering
  \includegraphics[width=0.49\linewidth]{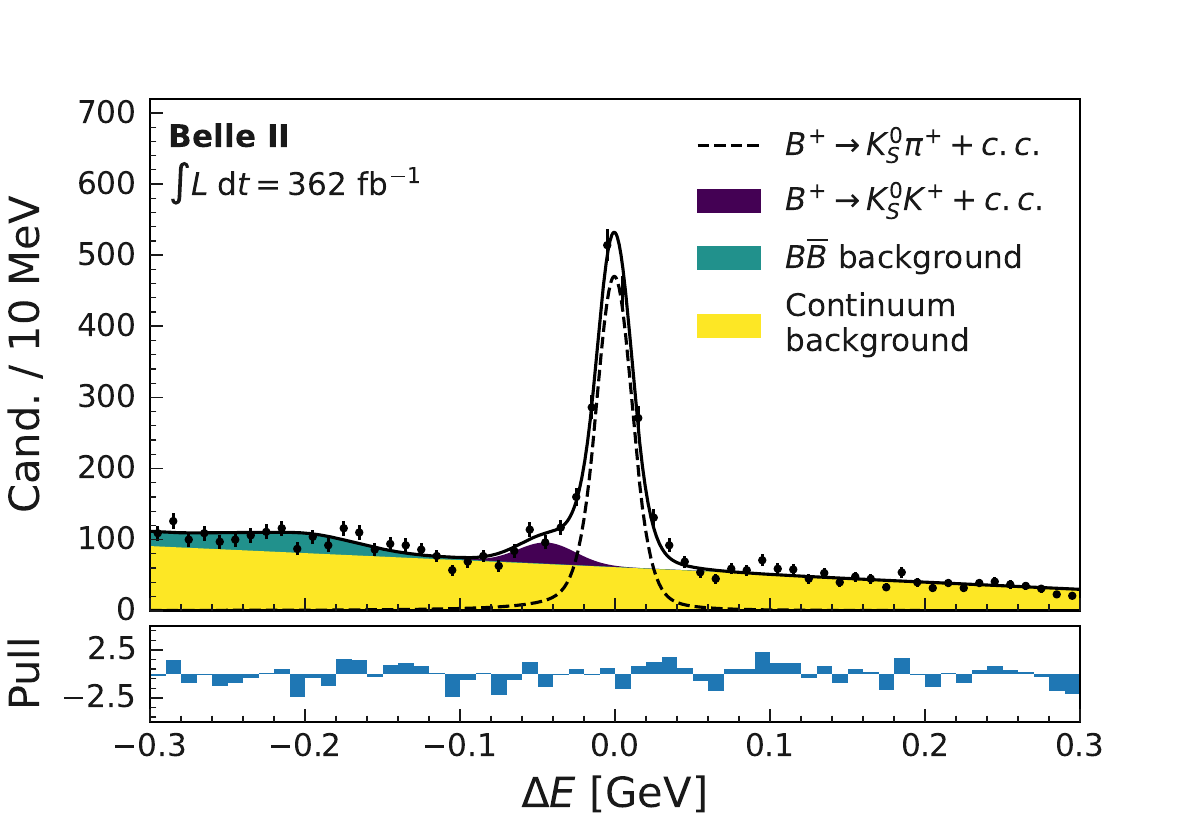}
  \includegraphics[width=0.49\linewidth]{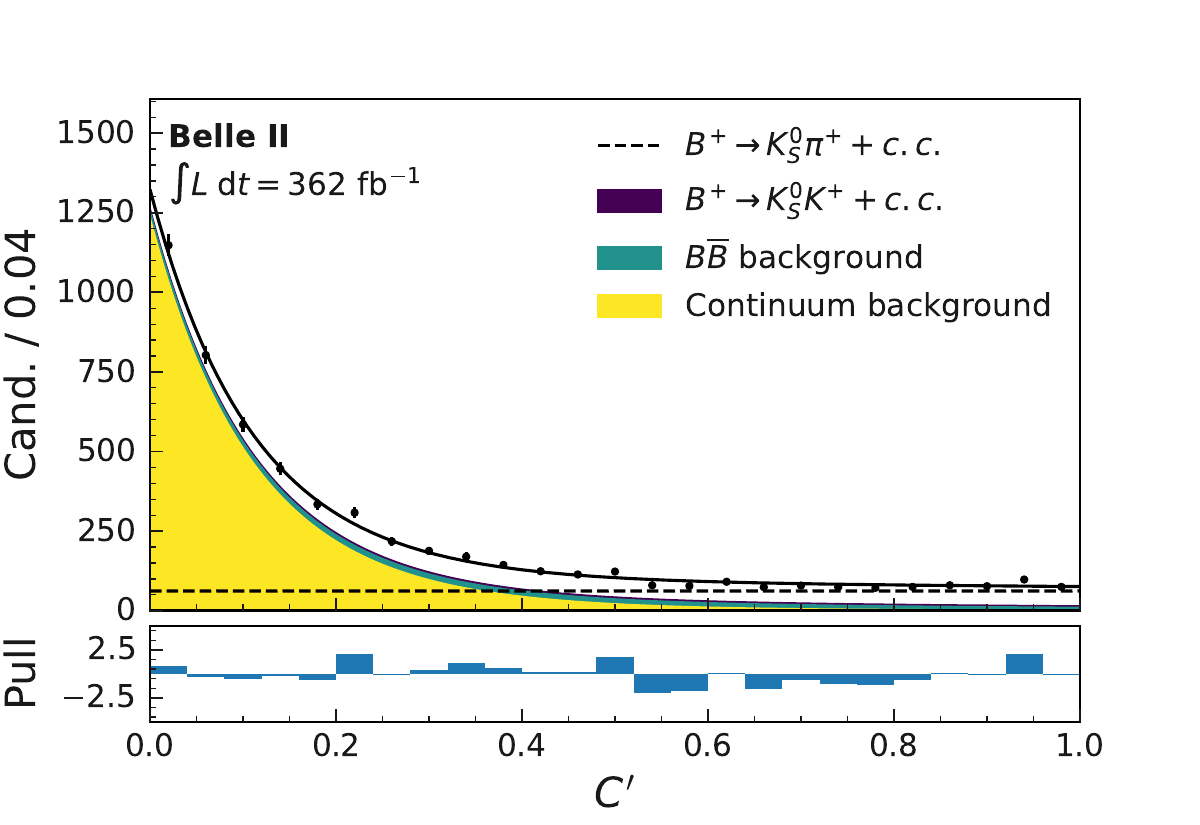}
  \includegraphics[width=0.49\linewidth]{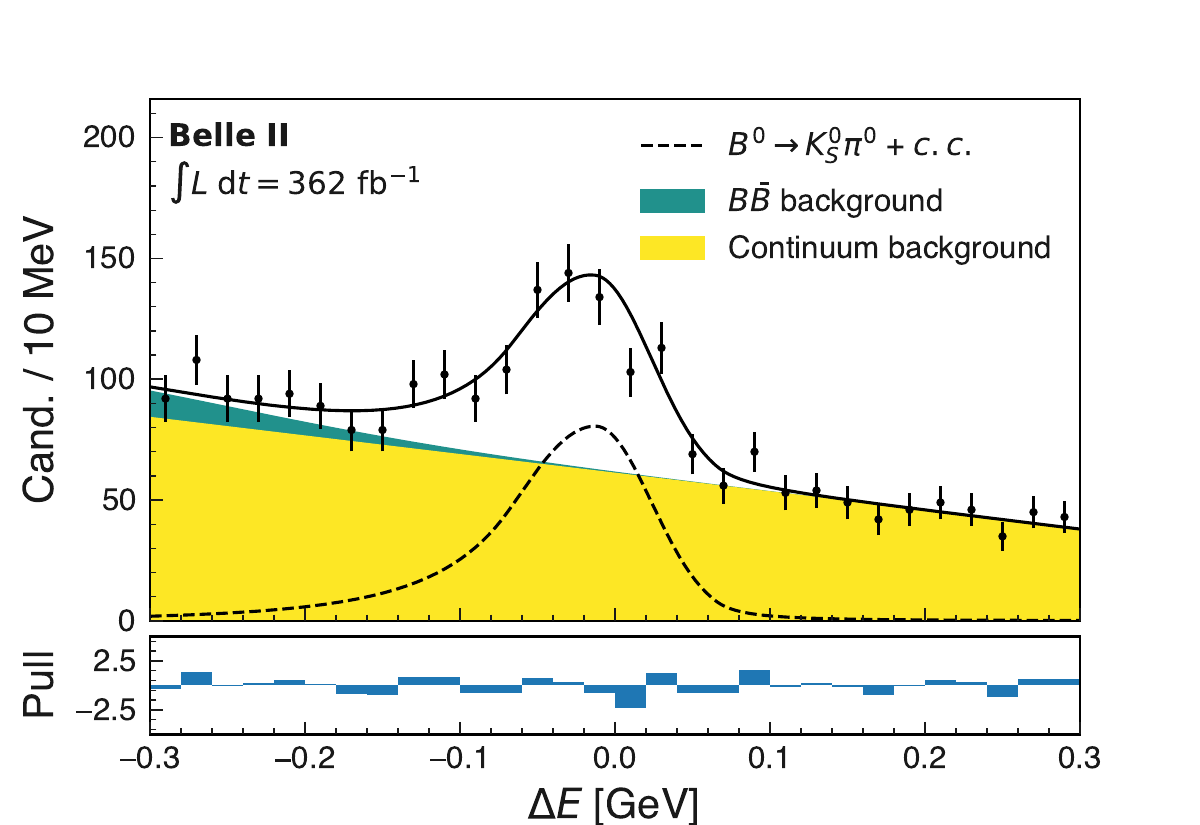}
  \includegraphics[width=0.49\linewidth]{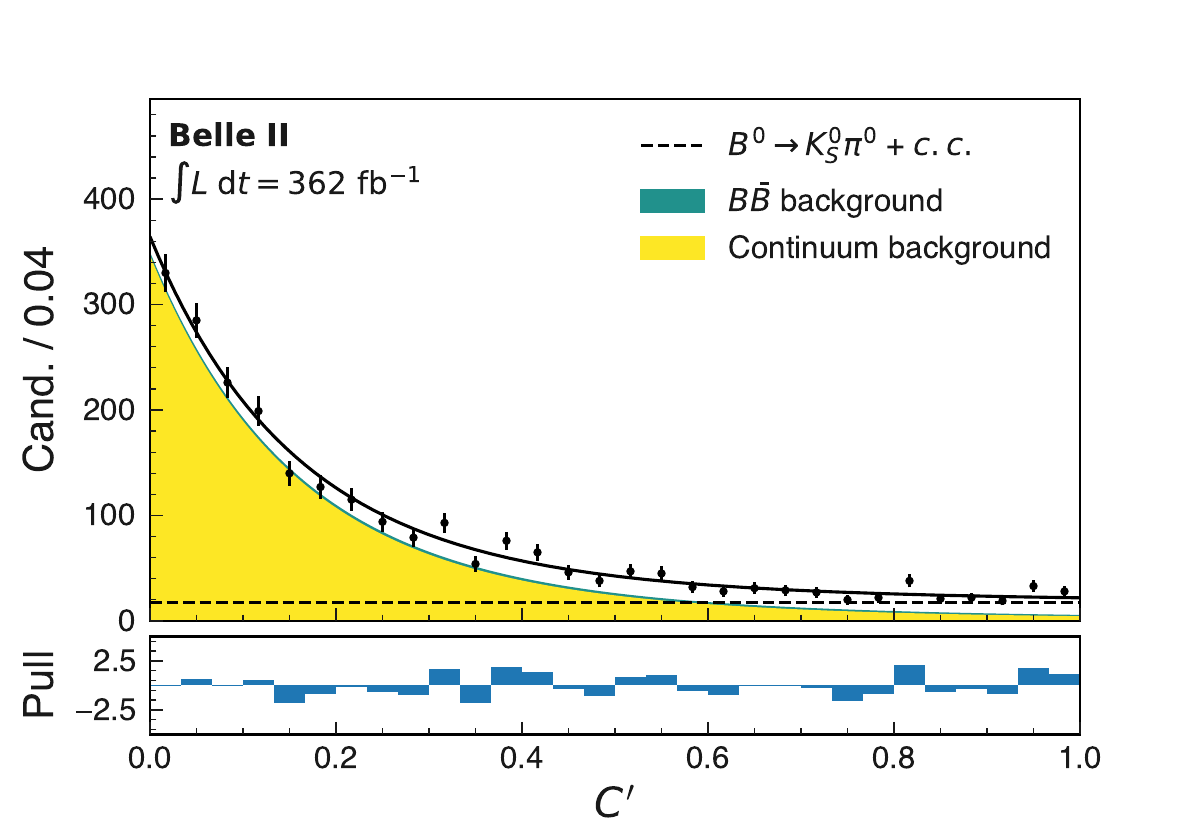}
  \caption{Distribution of (left) $\Delta E$ and (right) transformed CS output $C'$ for the (top) $\Bp \to \KS \pip$ and (bottom) $\Bz \to \KS \piz$ candidates with fit results overlaid. Points with error bars are data, the total fit is shown as a solid black curve, the signal as a black dashed curve, the peaking background as a purple shaded area, the $\BB$ background as a green shaded area, and the continuum background as a yellow shaded area. Fit pulls are also shown.
  }
  \label{fig:fit_result_3}
\end{figure*}

The fit results are listed in Table~\ref{tab:fit_result}. The fit projections onto $\Delta E$ and $C'$ are shown in Figs.~\ref{fig:fit_result} to \ref{fig:fit_result_3}.
All results agree with world averages.
We obtain a precision comparable to the best existing determinations reported in previous studies from $\epem$ \mbox{$B$-factory} experiments, Belle~\cite{PhysRevD.87.031103, Dalseno_2013, Fujikawa_2010} and \babar~\cite{Lees_2013, Aubert_2007_h+h-, Aubert_2007_h+pi0, Aubert_2009}.
When normalizing our statistical uncertainties to the same number of produced $B$ mesons, our precision surpasses that of Belle by up to 38\%.
This is due to data-driven CS and a better $\piz$ selection.
The measurement of the branching fraction of $\Bz \to \pip \pim$ is the most precise determination by a single experiment to date. 

The branching fraction and direct {\it CP} asymmetry of $\Bz \to \Kz \piz$ decays have also been measured in an analysis of the decay time evolution~\cite{sagr:kspi0}, which is based on the same data sample as this work but features a different event selection. 
The systematic uncertainty on the branching fraction given in Ref.~\cite{sagr:kspi0} was not reported previously; here we evaluate that uncertainty to be 5.6\% (relative) and update the measurement to be $\mathcal{B}_{K^0\piz} = (11.15 \pm 0.68 \pm 0.62)\times 10^{-6}$. We combine this updated branching fraction and the $\mathcal{A}_{{\it CP}}$ measured in Ref.~\cite{sagr:kspi0} with the results presented here, taking into account both statistical and systematic correlations. 
The systematic uncertainties are almost fully correlated between the two analyses. 
Statistically, the fraction of common candidates between the two analyses is 53\% in a signal-enhanced region defined by ${-0.13 < \Delta E <0.10\gev}$ and $C'>0.9$.
We assess the statistical correlation between the measurements by subdividing the data into three samples that contain (a) overlapping events, (b) events only found in this analysis, and (c) events only found in the analysis of the decay time evolution. 
We generate an ensemble of 1000 replicas of these samples by randomly selecting the events with replacement, allowing repetition of events.
We analyze the union of the (a) and (b) samples using the fitter presented in this paper, and the union of the (a) and (c) samples using the fitter of Ref.~\cite{sagr:kspi0}. 
By comparing the results of all members of the ensemble, we estimate the statistical correlation between the branching-fraction measurements from the two analyses to be 76\%, and that of the direct {\it CP} asymmetry to be 21\%.
The correlation is lower for the direct {\it CP} asymmetry as the time-dependent analysis gains additional sensitivity from the fit to the decay time.
The measurements reported here and those of Ref.~\cite{sagr:kspi0} are consistent; hence we combine them, using a linear unbiased estimator~\cite{BLUE}. 
The results are reported in the last row of Table~\ref{tab:fit_result}. The result for the direct {\it CP} asymmetry supersedes the measurement reported in Ref.~\cite{sagr:kspi0} and is the most precise determination by a single experiment to date.

\section{Systematic Uncertainties}
\label{sec:syst}

The main sources of systematic uncertainty are listed in Table~\ref{tab:syst_br} for the branching fractions and Table~\ref{tab:syst_acp} for the direct {\it CP} asymmetries. 

We assign an uncertainty on the branching fractions of 0.24\% for each track in the final state, to take into account tracking efficiency uncertainties, which are obtained from $\epem \to \tautau$ events, where one $\tau$ decays leptonically as $\taup \to \ellp \nul \nutb$ with $\ell = e, \mu$ and the other hadronically as $\taum \to \pim \pi^{\pm} \pi^{\mp} (N\piz)\nut$, where $N\geq0$. 
A 1.5\% systematic uncertainty is assigned to each branching fraction due to the uncertainty on the number $\mathcal{N}$ of $\BB$ pairs. 
In addition, the uncertainty on $f^{+-/00}$, 2.4\% for $\BpBm$ and 2.5\% for $\BzBzb$~\cite{f+-/f00}, is included as a systematic uncertainty.

For all corrections to the reconstruction efficiency described in Sec.~\ref{sec:corrections}, except for the PID correction, we assign the uncertainty of the correction as a systematic uncertainty on the branching fraction.
The largest systematic uncertainty comes from the $\piz$ correction, which is dominated by the uncertainty of the branching fraction ratio between $\Dz \to \Km \pip \piz$ and $\Dz \to \Km \pip$ (3.6\%)~\cite{PDG_2022}.
To estimate uncertainties associated with the PID corrections, we propagate the uncertainties using experiments simulated by drawing events from the PDF, with nominal and alternative corrections obtained by varying the PID corrections within their uncertainties.
We calculate the difference between the fit results using nominal and alternative corrections, and assign the standard deviation of the difference distribution as a systematic uncertainty.

Systematic uncertainties associated with the PDF correction factors, {\emph{i.e.}, the shift and scaling parameters, are assessed by repeating the fit on simulated event samples with nominal and alternative correction parameters drawn from the relevant Gaussian distributions, taking correlations into account. 
We calculate the difference between the fit results using nominal and alternative correction parameters, and assign the standard deviation of the difference distribution as a systematic uncertainty.
A similar procedure is applied for the signal, feed-across and $\KS\Kp$ peaking-background shapes.
To assess the systematic uncertainty associated with the $\BB$ background shapes, we construct alternative fit models. For the $\Delta E$ distribution, we employ alternative PDFs that yield comparable fit quality on simulated samples. For $C'$, we first verify, using control samples, that simulation accurately reproduces the $C'$ distribution of $B$ decays. We then vary the shape parameters, previously fixed from the fit to simulated samples, within their respective statistical uncertainties.
We generate simulated experiments by drawing events from the PDF based on this alternative model and fit them with the nominal and alternative fit model.  
We assign the average deviations of the fitted values of $
\mathcal{B}$ and $\mathcal{A}_{{\it CP}}$ between the alternative and nominal model as systematic uncertainties.

We use all candidates in events with multiple candidates. To assess a systematic uncertainty associated with a possible difference between data and simulation in candidate multiplicity, we repeat the fit to the data by randomly selecting a single candidate in each event. The difference in results from the default fit result is taken as a systematic uncertainty.

For $\Bz \to \KS \piz$, the $\BB$ and continuum backgrounds are assumed to be flavor-symmetric. 
To evaluate uncertainties from background asymmetries, we generate experiments using events drawn from background PDFs with an assumed asymmetry.  
For the continuum background we set the asymmetry equal to that observed for candidates with $0.1 < \Delta E < 0.3 \gev$ and assign the mean shift between the results obtained in these experiments and the nominal fit as a systematic uncertainty.  
For the $\BB$ background asymmetry we generate experiments with the asymmetry set to $\pm1$.  
The mean shift between the results obtained in these experiments and the nominal result is scaled by $1/\sqrt{3}$ to obtain a 68\% confidence level uncertainty.

In the fit to the $\Bz \to \KS\piz$ sample, the flavor-tagging parameters are Gaussian-constrained with widths corresponding to their uncertainties in the default fit; thus, any systematic uncertainty related to those parameters is absorbed in the statistical uncertainty.
The value of the decay-time-integrated $\BzBzb$ mixing probability $\chi_d$ is fixed in the fit. We propagate its uncertainty using simplified simulated experiments and find that it is negligible.
We estimate the systematic uncertainty of the continuum flavor parameters by generating simulated samples with the continuum flavor parameters equal to those observed for candidates with $0.1<\Delta E<0.3 \gev$, where only continuum background is expected.
We calculate the difference between the fit results using nominal and alternative continuum flavor parameters, and assign the standard deviations of the difference distributions as systematic uncertainties.
We assign the uncertainty of the instrumental asymmetry described in Sec.~\ref{sec:corrections} as a systematic uncertainty on $\mathcal{A}_{CP}$.

We study potential bias in our fit result by generating simulated event samples with various input values of $\mathcal{B}$ and $\mathcal{A}_{{\it CP}}$ and fitting these samples in the same manner as we fit the data. 
We observe a small biases for the {\it CP} asymmetries of $\Bp \to \KS \pip$ and $\Bp \to \pip \piz$, which are assigned as systematic uncertainties.

\begin{table*}
  \caption{Summary of the relative systematic uncertainties (\%) on the branching fractions.}
  \label{tab:syst_br}
  \centering
  \begin{tabular}{l ccc ccc}
  \hline
  \hline
  Source & $\Bz \to \Kp \pim$ & $\Bz \to \pip \pim$ & $\Bp \to \Kp \piz$ & $\Bp \to \pip \piz$ & $\Bp \to \KS \pip$ & $\Bz \to \KS \piz$\\
  \hline
  Tracking & 0.5 & 0.5 & 0.2 & 0.2 & 0.7 & 0.5\\
  $N_{\BB}$ & 1.5 & 1.5 & 1.5 & 1.5 & 1.5 & 1.5\\
  $f^{+-/00}$ & 2.5 & 2.5 & 2.4 & 2.4 & 2.4 & 2.5 \\
  $\piz$ efficiency &  &  & 3.8 & 3.8 &  & 3.8 \\
  $\KS$ efficiency &  &  &  &  & 2.0 & 2.0\\
  CS efficiency & 0.2 & 0.2 & 0.7 & 0.7 & 0.5 & 1.7\\
  PID correction & 0.1 & 0.1 & 0.1 & 0.2 &  & \\
  $\Delta E$ shift and scale & 0.1 & 0.2 & 1.2 & 2.0 & 0.3 & 1.7\\
  $K\pi$ signal model   & 0.1 & 0.2 & 0.1 & $<$0.1  & $<$0.1 & 0.1\\
  $\pi\pi$ signal model  & $<$0.1 & 0.1 & $<$0.1 & $<$0.1 &  &  \\
  $K\pi$ feed-across model     & $<$0.1 & 0.1 & $<$0.1 & 0.1 &  &  \\
  $\pi\pi$ feed-across model    & 0.1 & 0.2 & $<$0.1 & 0.1 &  &  \\
  $\KS\Kp$ model    &  &  &  &  & 0.1 &  \\
  \BB model &  &  & 0.3 & 0.5 & $<$0.1 & 0.3\\
  \qqbar flavor model &  &  &  &  &  & 0.9 \\
  Multiple candidates & $<$0.1 & $<$0.1 & 1.0 & 0.3 & 0.1 & 0.3\\
  \hline
  Total & 3.0 & 3.0 & 5.1 & 5.2 & 3.6 & 5.8 \\ 
  \hline
  \hline
  \end{tabular}
\end{table*}

\begin{table*}[]
  \caption{Summary of the absolute systematic uncertainties on the {\it CP} asymmetries.}
  \label{tab:syst_acp}
  \centering
  \begin{tabular}{l ccccc}
  \hline
  \hline
  Source & $\Bp \to \Kp \pim$ & $\Bp \to \Kp \piz$ & $\Bp \to \pip \piz$ & $\Bp \to \KS \pip$ & $\Bz \to \KS \piz$\\
  \hline
  $\Delta E$ shift and scale & $<$0.001 & 0.001 & 0.002 & 0.001 & 0.003\\
  $\KS\Kp$ model     &  &  &  & 0.001 & \\
  $\B\Bbar$ background asymmetry &  &  &  &  & 0.026\\
  $\q\qbar$ background asymmetry &  &  &  &  & 0.024 \\
  \qqbar flavor model &  &  &  &  & 0.011 \\
  Fitting bias &  &  & 0.007 & 0.006 &  \\
  Instrumental asymmetry & 0.007 & 0.005 & 0.004 & 0.004 &  \\
  \hline
  Total & 0.007 & 0.005 & 0.008 & 0.007 & 0.037\\
  \hline
  \hline
  \end{tabular}
\end{table*}

\section{Test of the sum rule and conclusion}
We test the sum rule of Eq.~(\ref{equ:isorule}) using our measurements of the branching fractions and {\it CP} asymmetries and the measured ratio $\tau_{\Bz}/\tau_{\Bp} = 0.9273 \pm 0.0033$~\cite{PDG_2022}. 
The ratios of branching fractions are summarized in Table~\ref{tab:ratios}.
Common systematic uncertainties, such as those related to the tracking efficiency and the number of produced $B$ mesons, divide out. 
The systematic uncertainty from $f^{00}$ also cancels in the ratio for $\Bz$ decays. We consider the anticorrelation of $f^{00}$ and $f^{+-}$ uncertainties for the ratio between $\Bp$ and $\Bz$ decays.
We obtain a value for the sum rule parameter of 
\begin{equation}
I_{K\pi} = -0.03 \pm 0.13 \pm 0.04,  
\end{equation}
accounting for correlations between uncertainties. 
This value is consistent with theoretical expectations.

To conclude, we report measurements of the branching fractions for $\Bz \to \Kp \pim$, $\Bz \to \pip \pim$, $\Bp \to \pip \piz$, $\Bp \to \Kp \piz$, $\Bp \to \Kz \pip$, and $\Bz \to \Kz \piz$ and the {\it CP} asymmetries for all modes apart from ${\Bz \to \pip \pim}$.
The results agree with current world averages~\cite{PDG_2022} and have precision comparable to the current best results despite using a smaller sample for most channels.
The measurements of the branching fraction of the $\Bz \to \pip \pim$ decay, and of the direct {\it CP} asymmetry (combined with Ref.~\cite{sagr:kspi0}) of the $\Bz \to \Kz \piz$ decay are the most precise determinations to date by a single experiment.

Using only Belle II measurements of branching fractions and asymmetries, we obtain a value of the sum rule in agreement with the SM expectation. 
Our precision is limited by sample size and is similar to the precision resulting from the average of measurements by the Belle, \babar, and LHCb collaborations~\cite{HFLAV}.

\begin{table}
  \centering
  \caption{Ratios of branching fractions used as input for the calculation of $I_{K\pi}$.}
  \label{tab:ratios}
  \begin{tabular}{l rcl cl}
  \hline
  \hline
    Modes & \multicolumn{5}{c}{Ratio} \\
  \hline
  ${\mathcal{B}_{K^0\pi^+}}/{\mathcal{B}_{K^+\pi^-}}$ & 1.180 & $\pm$ & 0.040 & $\pm$ & 0.064 \\ 
  ${\mathcal{B}_{K^+\pi^0}}/{\mathcal{B}_{K^+\pi^-}}$ & 0.674 & $\pm$ & 0.022 & $\pm$ & 0.044 \\ 
  ${\mathcal{B}_{K^0\pi^0}}/{\mathcal{B}_{K^+\pi^-}}$ & {0.519} & $\pm$ & {0.032} & $\pm$ & {0.026} \\ 
    \hline
    \hline
  \end{tabular}
\end{table}

\section*{Acknowledgements}

This work, based on data collected using the Belle II detector, which was built and commissioned prior to March 2019, was supported by
Science Committee of the Republic of Armenia Grant No.~20TTCG-1C010;
Australian Research Council and Research Grants
No.~DP200101792, 
No.~DP210101900, 
No.~DP210102831, 
No.~DE220100462, 
No.~LE210100098, 
and
No.~LE230100085; 
Austrian Federal Ministry of Education, Science and Research,
Austrian Science Fund
No.~P~31361-N36
and
No.~J4625-N,
and
Horizon 2020 ERC Starting Grant No.~947006 ``InterLeptons'';
Natural Sciences and Engineering Research Council of Canada, Compute Canada and CANARIE;
National Key R\&D Program of China under Contract No.~2022YFA1601903,
National Natural Science Foundation of China and Research Grants
No.~11575017,
No.~11761141009,
No.~11705209,
No.~11975076,
No.~12135005,
No.~12150004,
No.~12161141008,
and
No.~12175041,
and Shandong Provincial Natural Science Foundation Project~ZR2022JQ02;
the Czech Science Foundation Grant No.~22-18469S;
European Research Council, Seventh Framework PIEF-GA-2013-622527,
Horizon 2020 ERC-Advanced Grants No.~267104 and No.~884719,
Horizon 2020 ERC-Consolidator Grant No.~819127,
Horizon 2020 Marie Sklodowska-Curie Grant Agreement No.~700525 ``NIOBE''
and
No.~101026516,
and
Horizon 2020 Marie Sklodowska-Curie RISE project JENNIFER2 Grant Agreement No.~822070 (European grants);
L'Institut National de Physique Nucl\'{e}aire et de Physique des Particules (IN2P3) du CNRS
and
L'Agence Nationale de la Recherche (ANR) under grant ANR-21-CE31-0009 (France);
BMBF, DFG, HGF, MPG, and AvH Foundation (Germany);
Department of Atomic Energy under Project Identification No.~RTI 4002,
Department of Science and Technology,
and
UPES SEED funding programs
No.~UPES/R\&D-SEED-INFRA/17052023/01 and
No.~UPES/R\&D-SOE/20062022/06 (India);
Israel Science Foundation Grant No.~2476/17,
U.S.-Israel Binational Science Foundation Grant No.~2016113, and
Israel Ministry of Science Grant No.~3-16543;
Istituto Nazionale di Fisica Nucleare and the Research Grants BELLE2;
Japan Society for the Promotion of Science, Grant-in-Aid for Scientific Research Grants
No.~16H03968,
No.~16H03993,
No.~16H06492,
No.~16K05323,
No.~17H01133,
No.~17H05405,
No.~18K03621,
No.~18H03710,
No.~18H05226,
No.~19H00682, 
No.~22H00144,
No.~22K14056,
No.~22K21347,
No.~23H05433,
No.~26220706,
and
No.~26400255,
the National Institute of Informatics, and Science Information NETwork 5 (SINET5), 
and
the Ministry of Education, Culture, Sports, Science, and Technology (MEXT) of Japan;  
National Research Foundation (NRF) of Korea Grants
No.~2016R1\-D1A1B\-02012900,
No.~2018R1\-A2B\-3003643,
No.~2018R1\-A6A1A\-06024970,
No.~2019R1\-I1A3A\-01058933,
No.~2021R1\-A6A1A\-03043957,
No.~2021R1\-F1A\-1060423,
No.~2021R1\-F1A\-1064008,
No.~2022R1\-A2C\-1003993,
and
No.~RS-2022-00197659,
Radiation Science Research Institute,
Foreign Large-Size Research Facility Application Supporting project,
the Global Science Experimental Data Hub Center of the Korea Institute of Science and Technology Information
and
KREONET/GLORIAD;
Universiti Malaya RU grant, Akademi Sains Malaysia, and Ministry of Education Malaysia;
Frontiers of Science Program Contracts
No.~FOINS-296,
No.~CB-221329,
No.~CB-236394,
No.~CB-254409,
and
No.~CB-180023, and SEP-CINVESTAV Research Grant No.~237 (Mexico);
the Polish Ministry of Science and Higher Education and the National Science Center;
the Ministry of Science and Higher Education of the Russian Federation,
Agreement No.~14.W03.31.0026, and
the HSE University Basic Research Program, Moscow;
University of Tabuk Research Grants
No.~S-0256-1438 and No.~S-0280-1439 (Saudi Arabia);
Slovenian Research Agency and Research Grants
No.~J1-9124
and
No.~P1-0135;
Agencia Estatal de Investigacion, Spain
Grant No.~RYC2020-029875-I
and
Generalitat Valenciana, Spain
Grant No.~CIDEGENT/2018/020;
National Science and Technology Council,
and
Ministry of Education (Taiwan);
Thailand Center of Excellence in Physics;
TUBITAK ULAKBIM (Turkey);
National Research Foundation of Ukraine, Project No.~2020.02/0257,
and
Ministry of Education and Science of Ukraine;
the U.S. National Science Foundation and Research Grants
No.~PHY-1913789 
and
No.~PHY-2111604, 
and the U.S. Department of Energy and Research Awards
No.~DE-AC06-76RLO1830, 
No.~DE-SC0007983, 
No.~DE-SC0009824, 
No.~DE-SC0009973, 
No.~DE-SC0010007, 
No.~DE-SC0010073, 
No.~DE-SC0010118, 
No.~DE-SC0010504, 
No.~DE-SC0011784, 
No.~DE-SC0012704, 
No.~DE-SC0019230, 
No.~DE-SC0021274, 
No.~DE-SC0022350, 
No.~DE-SC0023470; 
and
the Vietnam Academy of Science and Technology (VAST) under Grant No.~DL0000.05/21-23.

These acknowledgements are not to be interpreted as an endorsement of any statement made
by any of our institutes, funding agencies, governments, or their representatives.

We thank the SuperKEKB team for delivering high-luminosity collisions;
the KEK cryogenics group for the efficient operation of the detector solenoid magnet;
the KEK computer group and the NII for on-site computing support and SINET6 network support;
and the raw-data centers at BNL, DESY, GridKa, IN2P3, INFN, and the University of Victoria for off-site computing support.

\bibliography{belle2}
\bibliographystyle{apsrev4-2}

\clearpage
\onecolumngrid

\end{document}